\documentstyle[12pt,prd,aps,epsfig,preprint]{revtex}
\tightenlines
\begin{document}
\draft
\date{\today}
\title{VMD, chiral loops, $\sigma$-meson, and $\omega-\rho$ mixing in $\omega\rightarrow \pi^0\pi^0\gamma$ decay}

\author{A. Gokalp~\thanks{agokalp@metu.edu.tr},  A. Kucukarslan and O. Yilmaz~\thanks{oyilmaz@metu.edu.tr}}
\address{ {\it Physics Department, Middle East Technical University,
06531 Ankara, Turkey}} \maketitle

\begin{abstract}
In an attempt to explain the latest experimental result about the
branching ratio of $\omega\rightarrow \pi^0\pi^0\gamma$ decay we
reexamine the decay mechanism of this decay in a phenomenological
framework in which the contributions of VMD, chiral loops,
$\sigma$-meson intermediate state amplitudes and the effects of
$\omega-\rho$ mixing are considered. We conclude that in order to
obtain the experimental value of the branching ratio
$B(\omega\rightarrow \pi^0\pi^0\gamma)$ $\sigma$-meson amplitude
which makes a substantial contribution should be included into the
reaction mechanism and the effects of $\omega-\rho$ mixing should
be taken into account. We also estimate the coupling constant
$g_{\omega\sigma\gamma}$ as $g_{\omega\sigma\gamma}=0.11$ which is
much smaller than the values suggested by light cone QCD sum rules
calculations.
\end{abstract}

\thispagestyle{empty} ~~~~\\ 
\newpage
\setcounter{page}{1}

The recent experimental study of $\rho\rightarrow
\pi^0\pi^0\gamma$ and $\omega\rightarrow \pi^0\pi^0\gamma$ decays
by SND Collaboration obtained the value
$B(\omega\rightarrow\pi^{0}\pi^{0}\gamma)=(6.6^{+1.4}_{-0.8}\pm
0.6)\times 10^{-5}$ for the branching ratio of the
$\omega\rightarrow \pi^0\pi^0\gamma$ decay  \cite{R1}. Their
result is in good agreement with GAMS Collaboration measurement of
$B(\omega\rightarrow\pi^{0}\pi^{0}\gamma)=(7.2\pm 2.5)\times
10^{-5}$ \cite{R2}, but it has a higher accuracy.

On the theoretical side, $\omega\rightarrow \pi^0\pi^0\gamma$
decay was first studied by Singer \cite{R3} who postulated that
this transition proceeds through the $\omega\rightarrow
(\rho)\pi^0\rightarrow\pi^0\pi^0\gamma$ mechanism involving
$\rho$-meson intermediate state. The contribution of intermediate
vector mesons (VMD) to the vector meson decays into two
pseudoscalars and a single photon $V\rightarrow PP'\gamma$ was
also considered by Bramon et al. \cite{R4} using standard
Lagrangians obeying the SU(3)-symmetry, and in particular for the
branching ratio of the decay $\omega\rightarrow \pi^0\pi^0\gamma$
they obtained the result
$B(\omega\rightarrow\pi^{0}\pi^{0}\gamma)=2.8\times 10^{-5}$. The
$V\rightarrow PP'\gamma$ decays have also been considered within
the framework of chiral effective Lagrangians using chiral
perturbation theory. Bramon et al. \cite{R5} studied various such
decays using this approach and they noted that if chiral
perturbation theory Lagrangians are used there is no tree-level
contribution to the amplitudes for the decay processes
$V\rightarrow PP'\gamma$, and moreover the one-loop contributions
are finite and to this order no counterterms are required. They
considered both $\pi\pi$ and $K\bar{K}$ intermediate loops. In the
good isospin limit $\pi$-loop contributions to $\omega\rightarrow
\pi^0\pi^0\gamma$ amplitude vanish, and the contribution of
K-loops is two orders of magnitude smaller than the contribution
of VMD amplitude. Therefore, the VMD amplitude essentially
accounts for the decay rate of the $\omega\rightarrow
\pi^0\pi^0\gamma$ decay. Guetta and Singer \cite{R6} recently
updated the theoretical value for the branching ratio
$B(\omega\rightarrow\pi^0\pi^0\gamma)$ of the decay
$\omega\rightarrow \pi^0\pi^0\gamma$ as
$B(\omega\rightarrow\pi^{0}\pi^{0}\gamma)=(4.1\pm 1.1)\times
10^{-5}$. In their calculation they noted that when the Born
amplitude for VMD mechanism is used the decay rate
$\omega\rightarrow\pi^{0}\pi^{0}\gamma$ is proportional to the
coupling constants $g_{\omega\rho\pi}^2$ and
$g_{\rho\pi\gamma}^2$, and they assumed that the decay
$\omega\rightarrow 3\pi$ proceeds with the same mechanism as
$\omega\rightarrow\pi^{0}\pi^{0}\gamma$, that is as
$\omega\rightarrow (\rho)\pi\rightarrow\pi\pi\pi$ \cite{R7}. They
use the experimental inputs for the decay rates
$\Gamma(\omega\rightarrow 3\pi)$,
$\Gamma(\rho^0\rightarrow\pi^0\gamma)$ and
$\Gamma(\rho\rightarrow\pi\pi)$, and furthermore they employ  a
momentum dependent width for $\rho$-meson. If a constant
$\rho$-meson width is used, then the value
$B(\omega\rightarrow\pi^{0}\pi^{0}\gamma)=(3.6\pm 0.9)\times
10^{-5}$ is obtained for the branching ratio of this decay.
Therefore there appears to be a serious discrepancy between the
theoretical result and the experimental value for the branching
ratio of the $\omega\rightarrow\pi^{0}\pi^{0}\gamma$ decay.

Guetta and Singer \cite{R6} noted that in the theoretical
framework based on chiral perturbation theory and vector meson
dominance one feature has been neglected. This is the possibility
of $\omega-\rho$ mixing one consequence of which is the isospin
violating $\omega\rightarrow\pi^{+}\pi^{-}$ decay with the
branching ratio B$(\omega\rightarrow\pi^{+}\pi^{-})=(2.21\pm
0.30)\%$ \cite{R8}. The phenomenon of $\omega-\rho$ mixing has
been observed in the electromagnetic form factor of the pion
improving the standard VMD model result involving $\rho$-meson
intermediate state. A recent review of $\omega-\rho$ mixing and
vector meson dominance is given by O`Connell et al. \cite{R9}.
Guetta and Singer \cite{R6} calculated the effect of $\omega-\rho$
mixing using the Born amplitude for VMD mechanism and showed that
it increases the $\omega\rightarrow\pi^{0}\pi^{0}\gamma$ width
only by $5\%$. They then combined all the improvements on the
simple Born term of VMD mechanism, that is $\omega-\rho$ mixing,
momentum dependence of $\rho$-meson width and the inclusion of the
chiral loop amplitude as given by Bramon et al. \cite{R5}, and
using the resulting amplitude which includes all these effects
they obtain the theoretical result
$B(\omega\rightarrow\pi^{0}\pi^{0}\gamma)=(4.6\pm 1.1)\times
10^{-5}$  for the branching ratio of the
$\omega\rightarrow\pi^{0}\pi^{0}\gamma$ decay. Palomar et al.
\cite{R10} also analyzed the radiative $V\rightarrow PP'\gamma$
decays using vector meson dominance, chiral loops obtained using
unitarized chiral perturbation theory, and $\omega-\rho$ mixing.
They obtained the result
$B(\omega\rightarrow\pi^{0}\pi^{0}\gamma)=(4.7\pm 0.9)\times
10^{-5}$  for this branching ratio. These theoretical results are
still seriously less than the latest experimental result
$B(\omega\rightarrow\pi^{0}\pi^{0}\gamma)=(6.6^{+1.4}_{-0.8}\pm
0.6)\times 10^{-5}$. Therefore, the possibility  of additional
contributions to the mechanism of
$\omega\rightarrow\pi^{0}\pi^{0}\gamma$ decay should be
investigated.

One such additional contribution to the
$\omega\rightarrow\pi^{0}\pi^{0}\gamma$ decay may be provided by
the amplitude involving scalar-isoscalar $\sigma$-meson as an
intermediate state. The existence of the controvertial
$\sigma$-meson now seems to be established by the Fermilab (E791)
Collaboration in their observation of the
$D^+\rightarrow\sigma\pi^+\rightarrow \pi\pi\pi$ decay channel in
which $\sigma$-meson is seen as a clear dominant peak with
$M_{\sigma}=(438\pm 31)$ MeV, and $\Gamma_\sigma=(338\pm 48)$ MeV
\cite{R11}, where statistical and systematic errors are added in
quadrature \cite{R12}. Two of present authors in a previous work
\cite{R13}, calculated the decay rate for the decay
$\omega\rightarrow\pi^{0}\pi^{0}\gamma$ by considering $\rho$-pole
vector meson dominance amplitude as well as $\sigma$-pole
amplitude in a phenomenological approach. By employing then the
available experimental value for this branching ratio
$B(\omega\rightarrow\pi^{0}\pi^{0}\gamma)=(7.2\pm 2.5)\times
10^{-5}$ \cite{R2} which is somewhat less accurate than the
present new value \cite{R1}, they obtained for the coupling
constant $g_{\omega\sigma\gamma}$ the values
$g_{\omega\sigma\gamma}=0.13$ and $g_{\omega\sigma\gamma}=-0.27$.
They observed that $\sigma$-meson intermediate state amplitude
makes an important contribution by itself and by its interference
with the VMD amplitude. The same authors later estimated the
coupling constant $g_{\omega\sigma\gamma}$  by studying
$\omega\sigma\gamma$-vertex in the light cone QCD sum rules method
\cite{R14}, and the value $\mid
g_{\omega\sigma\gamma}\mid=(0.72\pm 0.08)$ was deduced for this
coupling constant. Aliev et al. \cite{R15} also used light cone
QCD sum rules techniques to calculate the coupling constant
$g_{\rho\sigma\gamma}$, and they obtained the value
$g_{\rho\sigma\gamma}=(2.2\pm 0.2)$ from which by using
SU(3)-symmetry it follows that the coupling constant
$g_{\omega\sigma\gamma}$ should have the value
$g_{\omega\sigma\gamma}=0.73$. Thus, there seems to be a serious
discrepancy between the values obtained for the coupling constant
$g_{\omega\sigma\gamma}$ using light cone QCD sum rules method and
the phenomenological analysis of the
$\omega\rightarrow\pi^{0}\pi^{0}\gamma$ decay.

Therefore, in the present work, we reconsider the
$\omega\rightarrow\pi^{0}\pi^{0}\gamma$ decay in a
phenomenological framework in order to the assess the role of
$\sigma$-meson in the mechanism of
$\omega\rightarrow\pi^{0}\pi^{0}\gamma$ decay and to recalculate
the coupling constant $g_{\omega\sigma\gamma}$ utilizing the
latest experimental value of the branching ratio
$\omega\rightarrow\pi^{0}\pi^{0}\gamma$. For this purpose, we
calculate the decay rate for the decay
$\omega\rightarrow\pi^{0}\pi^{0}\gamma$ by considering $\rho$-pole
vector meson dominance amplitude, chiral loop amplitude,
$\sigma$-pole amplitude and we also include the effects of
$\omega-\rho$ mixing which was not taken into account in the
previous analysis \cite{R13}.

In order to calculate the effects of the $\omega-\rho$ mixing in
the $\omega\rightarrow\pi^{0}\pi^{0}\gamma$ decay we need an
amplitude characterizing the contribution of $\sigma$-meson to the
$\rho^0\rightarrow\pi^{0}\pi^{0}\gamma$ decay. In a previous work,
two of the present authors \cite{R16} calculated the branching
ratio $B(\rho^0\rightarrow\pi^{+}\pi^{-}\gamma)$ in a
phenomenological framework using pion bremsstrahlung amplitude and
$\sigma$-meson pole  amplitude, and they determined the coupling
constant $g_{\rho\sigma\gamma}$ by using the experimental value of
the branching ratio $B(\rho^0\rightarrow\pi^{+}\pi^{-}\gamma)$. In
a following work, these authors \cite{R17} calculated the
branching ratio $B(\rho^0\rightarrow\pi^{0}\pi^{0}\gamma)$ of the
$\rho^0\rightarrow\pi^{0}\pi^{0}\gamma$ decay using the value of
the coupling constant $g_{\rho\sigma\gamma}$ they thus obtained
again in a phenomenological approach in which the contribution of
$\sigma$-meson, $\omega$-meson intermediate states and of the
pion-loops are considered. However, the branching ratio
$B(\rho^0\rightarrow\pi^{0}\pi^{0}\gamma)$ obtained this way was
much larger than the experimental value. This unrealistic result
was due to the constant $\rho^0\rightarrow\sigma\gamma$ amplitude
employed and consequently the large coupling constant
$g_{\rho\sigma\gamma}$ that was deduced using the experimental
branching ratio of the $\rho^0\rightarrow\pi^{+}\pi^{-}\gamma$
decay. Therefore, it may be concluded that it is not realistic to
include $\sigma$-meson in the mechanism of radiative
$\rho^0$-meson decays as an intermediate pole state. On the other
hand, the $V\rightarrow PP'\gamma$ decays were also studied by
Marco et al. \cite{R18} in the framework of chiral perturbation
theory. They used the techniques of chiral unitary theory
\cite{R19}, and by a unitary resummation of the pion loops through
the Bethe-Salpeter equation they obtained the decay rates for
various decays. They furthermore noted that their result for
$\rho^0\rightarrow\pi^{0}\pi^{0}\gamma$ decay could be interpreted
as resulting from the $\rho^0\rightarrow
(\sigma)\gamma\rightarrow\pi^{0}\pi^{0}\gamma$ mechanism. Thus, it
seems that a natural way to include the effects of $\sigma$-meson
in the mechanism of radiative $\rho^0$-meson decays is to assume
that $\sigma$-meson couples to $\rho^0$-meson through the
pion-loop.

Our phenomenological approach is based on the Feynman diagrams
shown in Fig. 1 for $\omega\rightarrow\pi^0\pi^0\gamma$ decay and
in Fig. 2 for $\rho^{0}\rightarrow\pi^{0}\pi^{0}\gamma$ decay. The
direct terms shown in the diagrams in Fig. 1 b and in Fig. 2 b, c
are required to establish the gauge invariance. The interaction
term for two vector mesons and one pseudoscalar meson is given by
the Wess-Zumino anomaly term of the chiral Lagrangian \cite{R20},
we therefore  describe the $\omega\rho\pi$-vertex by the effective
Lagrangian
\begin{eqnarray} \label{e1}
{\cal L}^{eff}_{\omega\rho\pi}=g_{\omega\rho\pi}
\epsilon^{\mu\nu\alpha\beta}\partial_{\mu}\omega_{\nu}
\partial_{\alpha}\vec{\rho}_{\beta}\cdot\vec{\pi}~~,
\end{eqnarray}
which also conventionally defines the coupling constant
$g_{\omega\rho\pi}$. This coupling constant was determined by
Achasov et al. \cite{R21} through an experimental analysis as
$g_{\omega\rho\pi}=(14.4\pm 0.2)~~GeV^{-1}$, who assumed that
$\omega\rightarrow 3\pi$ decay proceeds with the intermediate
$\rho\pi$ state as
$\omega\rightarrow(\rho)\pi\rightarrow\pi\pi\pi$ and they used the
experimental value of the $\omega\rightarrow 3\pi$ width to deduce
the coupling constant $g_{\rho\omega\pi}$. Similarly we describe
the $V\pi\gamma$-vertices where $V=\rho,~\omega$ with the
effective Lagrangian
\begin{eqnarray}\label{e2}
{\cal L}^{eff}_{V\pi\gamma}=g_{V\pi\gamma}
\epsilon^{\mu\nu\alpha\beta}\partial_{\mu}V_{\nu}
\partial_{\alpha}A_{\beta}\pi~~.
\end{eqnarray}
We then use  the experimental partial widths  of the radiative
$V\rightarrow\pi\gamma$ decays \cite{R8} to deduce the coupling
constants $g_{\omega\pi\gamma}$  and $g_{\rho\pi\gamma}$. This way
for the coupling constants $g_{\omega\pi\gamma}$  and
$g_{\rho\pi\gamma}$ we obtaine the values
$g_{\omega\pi\gamma}=(0.706\pm 0.021)~~GeV^{-1}$ and
$g_{\rho\pi\gamma}=(0.274\pm 0.035)~~GeV^{-1}$. The
$\rho\pi\pi$-vertex is described by the effective Lagrangian
\begin{eqnarray}\label{e3}
{\cal L}^{eff}_{\rho\pi\pi}=g_{\rho\pi\pi}
\vec{\rho}_{\mu}\cdot(\partial^{\mu}\vec{\pi}\times\vec{\pi})~~,
\end{eqnarray}
and the experimental decay width of the decay
$\rho\rightarrow\pi\pi$ \cite{R8} then yields the value
$g_{\rho\pi\pi}=(6.03\pm 0.02)$ for the coupling constant
$g_{\rho\pi\pi}$.  We describe the  $\sigma\pi\pi$-vertex by the
effective Lagrangian
\begin{eqnarray} \label{e4}
{\cal L}^{eff}_{\sigma\pi\pi}=
\frac{1}{2}g_{\sigma\pi\pi}M_{\sigma}\vec{\pi}\cdot\vec{\pi}\sigma~~.
\end{eqnarray}
The experimental values for $M_{\sigma}$ and $\Gamma_\sigma$
$M_{\sigma}=(483\pm 31)$ MeV and $\Gamma_\sigma=(338\pm 48)$ MeV,
where statistical and systematic errors are added in quadrature
\cite{R11,R12} then results in the strong coupling constant
$g_{\sigma\pi\pi}=(5.3\pm 0.55)$. The effective Lagrangians ${\cal
L}^{eff}_{\sigma\pi\pi}$ and ${\cal L}^{eff}_{\rho\pi\pi}$ are
obtained from an extension of the $\sigma$ model where the
isovector $\rho$ is included through a Yang-Mills local gauge
theory based on isospin with the vector meson mass generated
through the Higgs mechanism \cite{R22}. In order to describe the
$\pi^{4}$-vertex we again consider the $\sigma$-model with
spontaneous symmetry breaking \cite{R23} and we describe the
$\pi^{4}$-vertex by the effective Lagrangian
\begin{eqnarray}\label{e5}
{\cal L}^{eff}=\frac{\lambda}{4}~(\vec{\pi}\cdot\vec{\pi})^{2}~~,
\end{eqnarray}
where the coupling constant $\lambda$ is given as
$\lambda=-\frac{g^2_{\pi NN}}{2}\frac{M_\sigma^2-M_\pi^2}{M_N^2}$
and we use $\frac{g_{\pi NN}^2}{4\pi}=14$. We note that this
effective interaction results in only isospin I=0 amplitudes. The
small I=2 amplitudes were also neglected in previous calculations
within the framework of chiral unitary theory \cite{R18}. The
$\omega\sigma\gamma$-vertex is described by the effective
Lagrangian
\begin{eqnarray}\label{e6}
{\cal
L}^{eff}_{\omega\sigma\gamma}=\frac{e}{M_\omega}g_{\omega\sigma\gamma}
\partial^\alpha\omega^\beta(\partial_{\alpha}A_\beta-\partial_{\beta}A_\alpha)\sigma~~,
\end{eqnarray}
which also defines the coupling constant $g_{\omega\sigma\gamma}$.

In our calculation of the invariant amplitude, we make the
replacement $q^2-M^2\rightarrow q^2-M^2+iM\Gamma$ in $\rho$-meson
and $\sigma$-meson propagators. We use for $\sigma$-meson the
momentum dependent width that follows from Eq. 4
\begin{eqnarray}\label{e7}
\Gamma_\sigma (q^2)=\Gamma_\sigma
\frac{M_\sigma^2}{q^2}\sqrt{\frac{q^2-4M_\pi^2}{M_\sigma^2-4M_\pi^2}}
\theta(q^2-4M_\pi^2)~~,
\end{eqnarray}
and for $\rho$-meson we use the following momentum dependent width
as conventionally adopted \cite{R9}
\begin{eqnarray}\label{e8}
\Gamma_\rho (q^2)=\Gamma_\rho
\frac{M_\rho}{\sqrt{q^2}}\left(\frac{q^2-4M_\pi^2}{M_\rho^2-4M_\pi^2}\right)^{3/2}
\theta(q^2-4M_\pi^2)~~.
\end{eqnarray}

Loop integrals similar to the ones appearing in Figs. 1 and 2 were
evaluated by Lucio and Pestiau \cite{R24} using dimensional
regularization. We use their results and, for example, we express
the contribution of the pion-loop amplitude corresponding to
$\rho^0\rightarrow(\pi^+\pi^-)\gamma\rightarrow\pi^0\pi^0\gamma$
reaction in Fig. 2 b as
\begin{eqnarray}\label{e9}
{\cal
M_\pi}=-\frac{e~g_{\rho\pi\pi}\lambda}{2\pi^{2}M_{\pi}^{2}}I(a,b)
\left[(p\cdot k)(\epsilon\cdot u)-(p\cdot\epsilon)(k\cdot
u)\right]~~,
\end{eqnarray}
where $a=M_\rho^2/M_\pi^2$, $b=(p-k)^2/M_\pi^{2}$, $p(u)$ and
$k(\epsilon)$ being the momentum (polarization vector) of
$\rho$-meson and photon, respectively. A similar amplitude
corresponding to
$\rho^0\rightarrow(\pi^+\pi^-)\gamma\sigma\rightarrow\gamma\pi^0\pi^0$
reaction can also be written. The function I(a,b) is given as
\begin{eqnarray}\label{e10}
I(a,b)=\frac{1}{2(a-b)} -\frac{2}{(a-b)^{2}}\left [
f\left(\frac{1}{b}\right)-f\left(\frac{1}{a}\right)\right ]
+\frac{a}{(a-b)^{2}}\left [
g\left(\frac{1}{b}\right)-g\left(\frac{1}{a}\right)\right ]
\end{eqnarray}
where
\begin{eqnarray}\label{e11}
&&f(x)=\left \{
\begin{array}{rr}
           -\left [ \arcsin (\frac{1}{2\sqrt{x}})\right ]^{2}~,& ~~x>\frac{1}{4} \\
\frac{1}{4}\left [ \ln (\frac{\eta_{+}}{\eta_{-}})-i\pi\right
]^{2}~, & ~~x<\frac{1}{4}
            \end{array} \right.
\nonumber \\ && \nonumber \\ &&g(x)=\left \{ \begin{array}{rr}
        (4x-1)^{\frac{1}{2}} \arcsin(\frac{1}{2\sqrt{x}})~, & ~~ x>\frac{1}{4} \\
 \frac{1}{2}(1-4x)^{\frac{1}{2}}\left [\ln (\frac{\eta_{+}}{\eta_{-}})-i\pi \right ]~, & ~~ x<\frac{1}{4}
            \end{array} \right.
\nonumber \\ && \nonumber \\ &&\eta_{\pm}=\frac{1}{2x}\left [
1\pm(1-4x)^{\frac{1}{2}}\right ] ~.
\end{eqnarray}

Although the contribution of chiral kaon-loop diagram shown in
Fig. 1 b to the decay rate of
$\omega\rightarrow\pi^{0}\pi^{0}\gamma$ decay is  small, we also
include the corresponding amplitude of this diagram in our
calculation for completeness. However, since we lack any
experimental information to describe the $\omega K^+K^-$-vertex
and  $K^+K^-$-$\pi^0\pi^0$ amplitude, for the contribution of this
diagram we use the amplitude given by Bramon et al. \cite{R5}
derived using chiral perturbation theory. This may not be entirely
consistent with the philosophy of our phenomenological approach,
but since its contribution is shown to be small \cite{R5} we do
not think that this way of including kaon-loop diagram into our
calculation constitutes a serious inconsistency. Moreover, in Fig.
2 b in addition to pion-loop intermediate state there is also a
contribution to $\rho^0\rightarrow\pi^{0}\pi^{0}\gamma$ decay
coming from $K\bar{K}$ intermediate state. However, as shown by
Bramon et al. \cite{R5}, these kaon-loop intermediate states give
a contribution to $\rho^0\rightarrow\pi^{0}\pi^{0}\gamma$ decay
which is $10^3$ times smaller than the contribution coming from
the charged-pion loops. Therefore, in our calculation we do not
take the kaon-loop amplitude in
$\rho^0\rightarrow\pi^{0}\pi^{0}\gamma$ decay into account.

We describe the $\omega-\rho$ mixing by an effective Lagrangian of
the form
\begin{eqnarray}\label{e12}
{\cal L}^{eff}_{\rho-\omega}=\Pi_{\rho\omega}\omega_\mu\rho^\mu~~,
\end{eqnarray}
where $\omega_\mu$ and $\rho_\mu$ denote pure isospin field
combinations. The corresponding physical states can therefore be
written as \cite{R9}
\begin{eqnarray}\label{e13}
\mid \rho >&=&\mid\rho,I=1>+\epsilon\mid\omega,I=0> \nonumber \\
\mid \omega >&=&\mid\omega,I=0>-\epsilon\mid\rho,I=1>~~,
\end{eqnarray}
where
\begin{eqnarray}\label{e14}
\epsilon=\frac{\Pi_{\rho\omega}}{M_\omega^2-M_\rho^2+iM_\rho\Gamma_\rho-iM_\omega\Gamma_\omega}~~.
\end{eqnarray}
O`Connell et al. \cite{R9} determined $\Pi_{\rho\omega}$ from fits
to $e^+e^-\rightarrow \pi^+\pi^-$ data as
$\Pi_{\rho\omega}=(-3800\pm 370)~~ MeV^2$. Then, using the
experimental values for $M_V$ and $\Gamma_V$, the mixing parameter
$\epsilon$ is obtained as $\epsilon=(-0.006+i 0.036)$. Another
effect of $\omega-\rho$ mixing besides the mixing of the states is
that it modifies the $\rho$-propagator in diagrams in Fig. 1 a as
\begin{eqnarray}\label{e15}
\frac{1}{D_\rho(s)}\rightarrow
\frac{1}{D_\rho(s)}\left(1+\frac{g_{\omega\pi\gamma}}{g_{\rho\pi\gamma}}\frac{\Pi_{\rho\omega}}{D_\rho(s)}\right)
\end{eqnarray}
where $D_\rho(s)=s-M_\rho^2+iM_\rho\Gamma_\rho(s)$. The amplitude
of the decay $\omega\rightarrow\pi^{0}\pi^{0}\gamma$ can then be
written as ${\cal M}={\cal M}_0+\epsilon {\cal M}'$ where ${\cal
M}_0$ includes the contributions coming from the diagrams shown in
Fig. 1 for $\omega\rightarrow\pi^{0}\pi^{0}\gamma$ and ${\cal M}'$
represents the contributions of the diagrams in Fig. 2 for
$\rho^0\rightarrow\pi^{0}\pi^{0}\gamma$.

We calculate the invariant amplitude ${\cal M}$(E$_{\gamma}$,
E$_{1}$) this way for the decay
$\omega\rightarrow\pi^{0}\pi^{0}\gamma$ from the corresponding
Feynman diagrams shown in Fig. 1 and 2 for the decays
$\omega\rightarrow\pi^0\pi^0\gamma$ and
$\rho^{0}\rightarrow\pi^{0}\pi^{0}\gamma$, respectively.The
differential decay probability for an unpolarized $\omega$-meson
at rest is then given as
\begin{eqnarray}\label{e16}
\frac{d\Gamma}{dE_{\gamma}dE_{1}}=\frac{1}{(2\pi)^{3}}~\frac{1}{8M_{\omega}}~
\mid {\cal M}\mid^{2} ,
\end{eqnarray}
where E$_{\gamma}$ and E$_{1}$ are the photon and pion energies
respectively. We perform an average over the spin states of the
vector meson and a sum over the polarization states of the photon.
The decay width is then obtained by integration
\begin{eqnarray}\label{e17}
\Gamma=\left(\frac{1}{2}\right)\int_{E_{\gamma,min.}}^{E_{\gamma,max.}}dE_{\gamma}
       \int_{E_{1,min.}}^{E_{1,max.}}dE_{1}\frac{d\Gamma}{dE_{\gamma}dE_{1}}
\end{eqnarray}
where the factor ($\frac{1}{2}$) is included because of the
$\pi^{0}\pi^{0}$ in the final state. The minimum photon energy is
E$_{\gamma, min.}=0$ and the maximum photon energy is given as
$E_{\gamma,max.}=(M_{\omega}^{2}-4M_{\pi}^{2})/2M_{\omega}$=341
MeV. The maximum and minimum values for pion energy E$_{1}$ are
given by
\begin{eqnarray}\label{e18}
\frac{1}{2(2E_{\gamma}M_{\omega}-M_{\omega}^{2})} [
-2E_{\gamma}^{2}M_{\omega}+3E_{\gamma}M_{\omega}^{2}-M_{\omega}^{3}
 ~~~~~~~~~~~~~~~~~~~~~~~~~~~~ \nonumber \\
\pm  E_{\gamma}\sqrt{(-2E_{\gamma}M_{\omega}+M_{\omega}^{2})
       (-2E_{\gamma}M_{\omega}+M_{\omega}^{2}-4M_{\pi}^{2})}~] ~.
\nonumber
\end{eqnarray}

The theoretical decay rate for
$\omega\rightarrow\pi^{0}\pi^{0}\gamma$ decay that we calculate
results in a quadric equation for the coupling constant
$g_{\omega\sigma\gamma}$, and using the experimental value for
this decay rate \cite{R1}, we obtain the values
$g_{\omega\sigma\gamma}=(0.11\pm 0.01) $ and
$g_{\omega\sigma\gamma}=(-0.21\pm 0.02) $ for the coupling
constant $g_{\omega\sigma\gamma}$. These values are smaller than
the values $g_{\omega\sigma\gamma}=0.13$ and
$g_{\omega\sigma\gamma}=-0.27$ that were obtained in the previous
phenomenological analysis which did not include the effects of
$\omega-\rho$ mixing \cite{R12}. We may therefore conclude that
$\omega-\rho$ mixing does make a reasonably substantial
contribution to the $\omega\rightarrow\pi^{0}\pi^{0}\gamma$ decay
amplitude when $\sigma$-meson intermediate state is taken into
account which then results in a reduced value for the coupling
constant $g_{\omega\sigma\gamma}$.

The resulting $\pi^0\pi^0$-invariant mass distribution for the
$\omega\rightarrow\pi^{0}\pi^{0}\gamma$ decay if we use the
coupling constant $g_{\omega\sigma\gamma}=0.11$ is plotted in Fig.
3 where we also indicate the contributions coming from the
different amplitudes. The interference term between the different
amplitudes is positive over the whole region. The contribution of
the VMD amplitude calculated from the diagrams in Fig. 1 a does
not change appreciably if we take into account the effect of
$\omega-\rho$ mixing by including the contribution coming from the
diagrams in Fig. 2 a The situation changes somewhat if we consider
VMD and chiral loop amplitudes and $\omega-\rho$ mixing as well.
However, the significant modification is obtained when we include
VMD, chiral loop, and $\sigma$-meson intermediate state amplitudes
with $\omega-\rho$ mixing.

Although the shapes of the various curves are quite similar, as it
can also be seen the corresponding branching ratios are
considerably different. If we use only the VMD amplitudes we
obtain for the branching ratio
$B(\omega\rightarrow\pi^{0}\pi^{0}\gamma)$ the values
$B(\omega\rightarrow\pi^{0}\pi^{0}\gamma)=3.96 \times 10^{-5}$ and
$B(\omega\rightarrow\pi^{0}\pi^{0}\gamma)=4.22 \times 10^{-5}$
without and with the effect of the $\omega-\rho$ mixing,
respectively. These results are in agreement with the calculation
of Guetta and Singer \cite{R6}. When we consider VMD amplitudes
and chiral loop amplitudes, the resulting values for the branching
ratio are $B(\omega\rightarrow\pi^{0}\pi^{0}\gamma)=3.98\times
10^{-5}$ and $B(\omega\rightarrow\pi^{0}\pi^{0}\gamma)=4.67\times
10^{-5}$ without and with the effect of the $\omega-\rho$ mixing
included, respectively. Again these calculated values are
consistent with the previous results, in particular, with the
results of Bramon et al. \cite{R5} and Palomar et al. \cite{R10}.
However, the main difference with previous results arises when we
consider the contribution of $\sigma$-meson intermediate state.
Indeed, the branching ratio
$B(\omega\rightarrow\pi^{0}\pi^{0}\gamma)$ that we calculate using
the full amplitude including the contributions of VMD, chiral
loop, and $\sigma$-meson intermediate state diagrams which is
$B(\omega\rightarrow\pi^{0}\pi^{0}\gamma)=7.29\times 10^{-5}$ is
reduced to the value
$B(\omega\rightarrow\pi^{0}\pi^{0}\gamma)=6.6\times 10^{-5}$ when
the effect of the $\omega-\rho$ mixing is included. We thus
observe that in this case the effect of $\omega-\rho$ mixing on
the amplitude of the $\omega\rightarrow\pi^{0}\pi^{0}\gamma$ decay
is reasonably pronounced and moreover the contribution of the
$\sigma$-meson intermediate state is quite substantial. We may
therefore conclude that $\sigma$-meson intermediate state and
$\omega-\rho$ mixing should be included in the analysis of
$\omega\rightarrow\pi^{0}\pi^{0}\gamma$ decay in order to explain
the latest experimental result \cite{R1}.

In Fig. 4, we plot the $\pi^0\pi^0$-invariant mass distribution we
obtain if we use the coupling constant
$g_{\omega\sigma\gamma}=-0.21$. In this case also the
$\sigma$-meson intermediate state amplitude makes a very
significant contribution, but the interference term resulting from
the $\sigma$-meson and VMD and chiral loop amplitudes is negative
over some regions of the spectrum. Moreover, the overall shape is
quite different from the previous case.

In our work, we consider a new approach to the mechanism of
$\rho^0\rightarrow\pi^{0}\pi^{0}\gamma$ decay. We assume that the
$\sigma$-meson couples to $\rho^0$-meson through the pion-loop and
thus we neglect a direct $\rho\sigma\gamma$-vertex. The assumed
mechanism of $\rho^0\rightarrow\pi^{0}\pi^{0}\gamma$ decay is
shown by the Feynman diagrams in Fig. 2. We can therefore
calculate the branching ratio of
$\rho^0\rightarrow\pi^{0}\pi^{0}\gamma$ decay based on these
Feynman diagrams. This way we obtain for the branching ratio of
$\rho^0\rightarrow\pi^{0}\pi^{0}\gamma$ decay the value
$B(\rho^0\rightarrow\pi^{0}\pi^{0}\gamma)=(3.81\pm 0.63)\times
10^{-5}$ which is in a good agreement the latest experimental
result
$B(\rho^0\rightarrow\pi^{0}\pi^{0}\gamma)=(4.1^{+1.0}_{-0.9}\pm
0.3)\times 10^{-5}$ \cite{R1}. Scalar $\sigma$-meson effects in
radiative $\rho^0$-meson decays are studied in a recent work by
two of the present authors in detail \cite{R25}. In that work, the
standard $\pi^+\pi^-\rightarrow \pi^0\pi^0$ amplitude of chiral
perturbation theory is used in the loop diagrams, and the value
$B(\rho^0\rightarrow\pi^{0}\pi^{0}\gamma)=(4.95\pm 0.82)\times
10^{-5}$ is obtained for the branching ratio. If we use the same
amplitude in the present work we then obtain the value
$B(\omega\rightarrow\pi^{0}\pi^{0}\gamma)=6.12 \times 10^{-5}$
using the full amplitude including the contributions of VMD,
chiral loop, and $\sigma$-meson intermediate state diagram as well
as the effects of $\omega-\rho$ mixing, which is not very
different from $B(\omega\rightarrow\pi^{0}\pi^{0}\gamma)=6.6\times
10^{-5}$ that is obtained employing the effective Lagrangian given
in Eq. 5 to describe the $\pi^4$-vertex.

An essential assumption of our work, therefore, is that there is
no SU(3) vector meson-sigma-gamma vertex. Thus, the
$\omega\sigma\gamma$-vertex cannot be related to the
$\rho\sigma\gamma$-vertex. The $\omega\sigma\gamma$-vertex that we
use may be considered as representing the effective final state
interactions in the $\pi\pi$-channel. The small value of the
coupling constant $g_{\omega\sigma\gamma}$ that we obtain leads to
a change in the Born amplitude of the
$\omega\rightarrow\pi^{0}\pi^{0}\gamma$ decay which is of the same
of the magnitude, as it is typical of final state interactions. As
a matter of fact, Levy and Singer \cite{R26} in a study of the
$\omega\rightarrow\pi^{0}\pi^{0}\gamma$ decay using
dispersion-theoretical approach shown that final state
interactions resulting in a decay rate of the same order of
magnitude as the one calculated from the Born term can be
parameterized with the effective-pole approximation.

As a final comment, we like to mention that our analysis suggests
that the coupling constant $g_{\omega\sigma\gamma}$ has actually a
much smaller value than obtained by light cone QCD sum rules
calculations.

\newpage
\begin{figure}\vspace*{1.0cm}\hspace{2.5cm}
\epsfig{figure=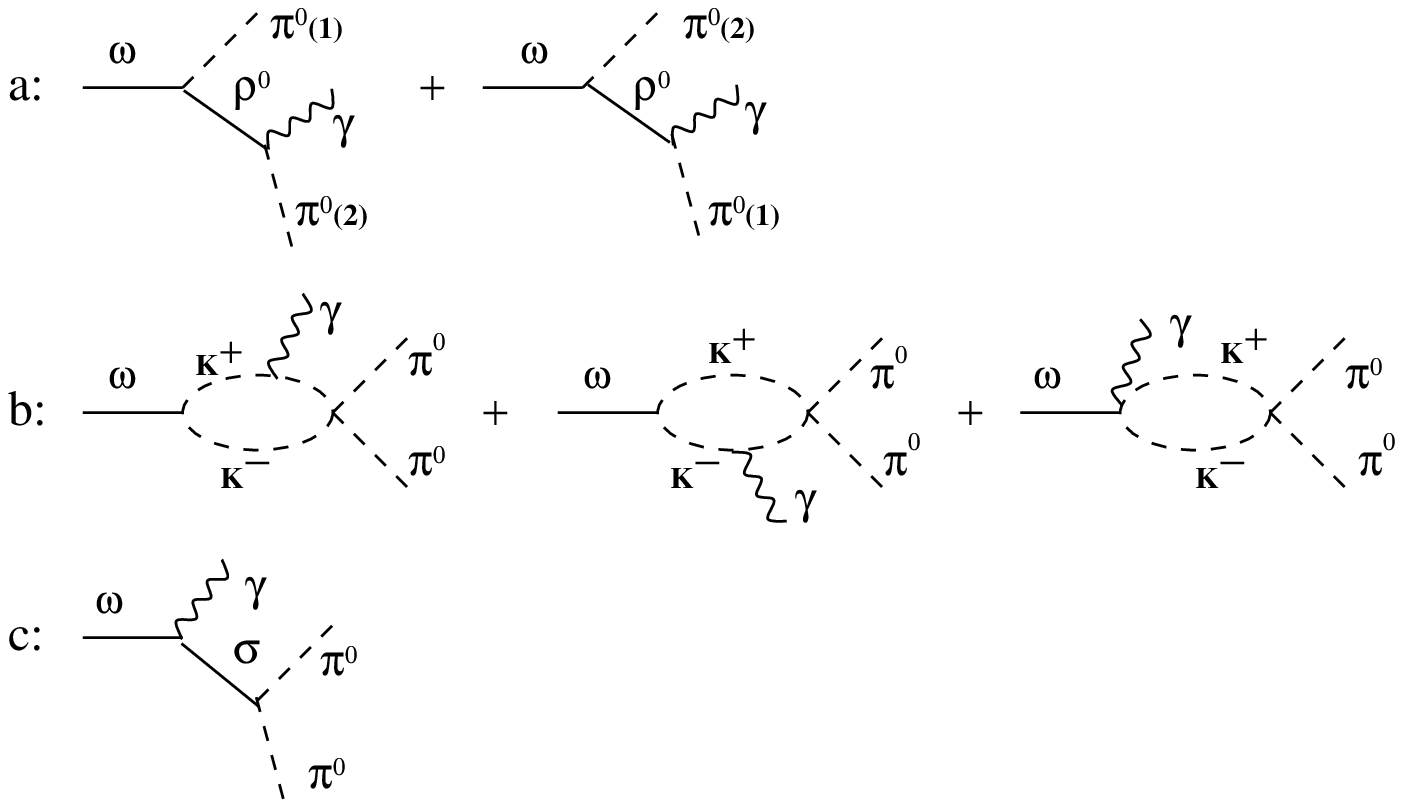,width=12cm,height=7cm,
angle=0}\vspace*{1.0cm} \caption{Feynman diagrams for the decay
$\omega\rightarrow\pi^0\pi^0\gamma$.} \label{fig1}
\end{figure}

\begin{figure}
\vspace*{1.0cm}\hspace{2.5cm}
\epsfig{figure=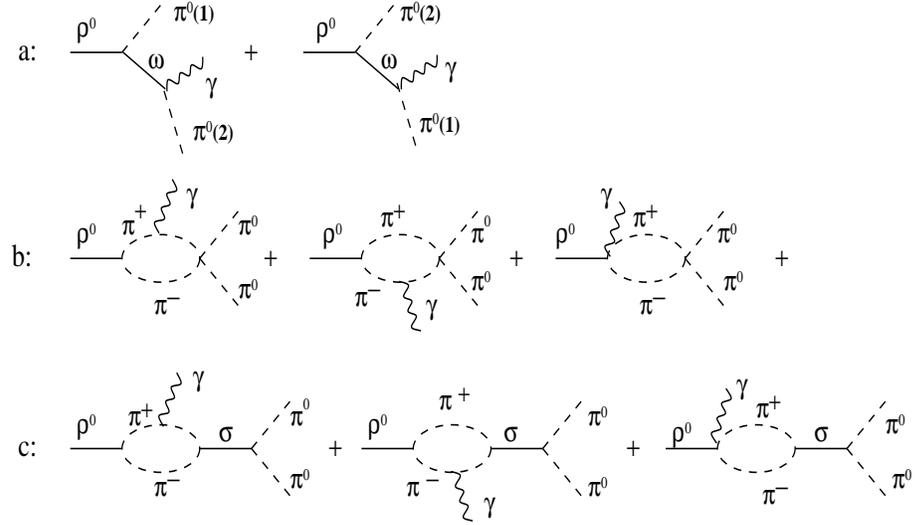,width=12cm,height=7cm,
angle=0}\vspace*{1.0cm} \caption{Feynman diagrams for the decay
$\rho^0\rightarrow\pi^0\pi^0\gamma$.} \label{fig2}
\end{figure}

\newpage
\begin{figure}
\epsfig{figure=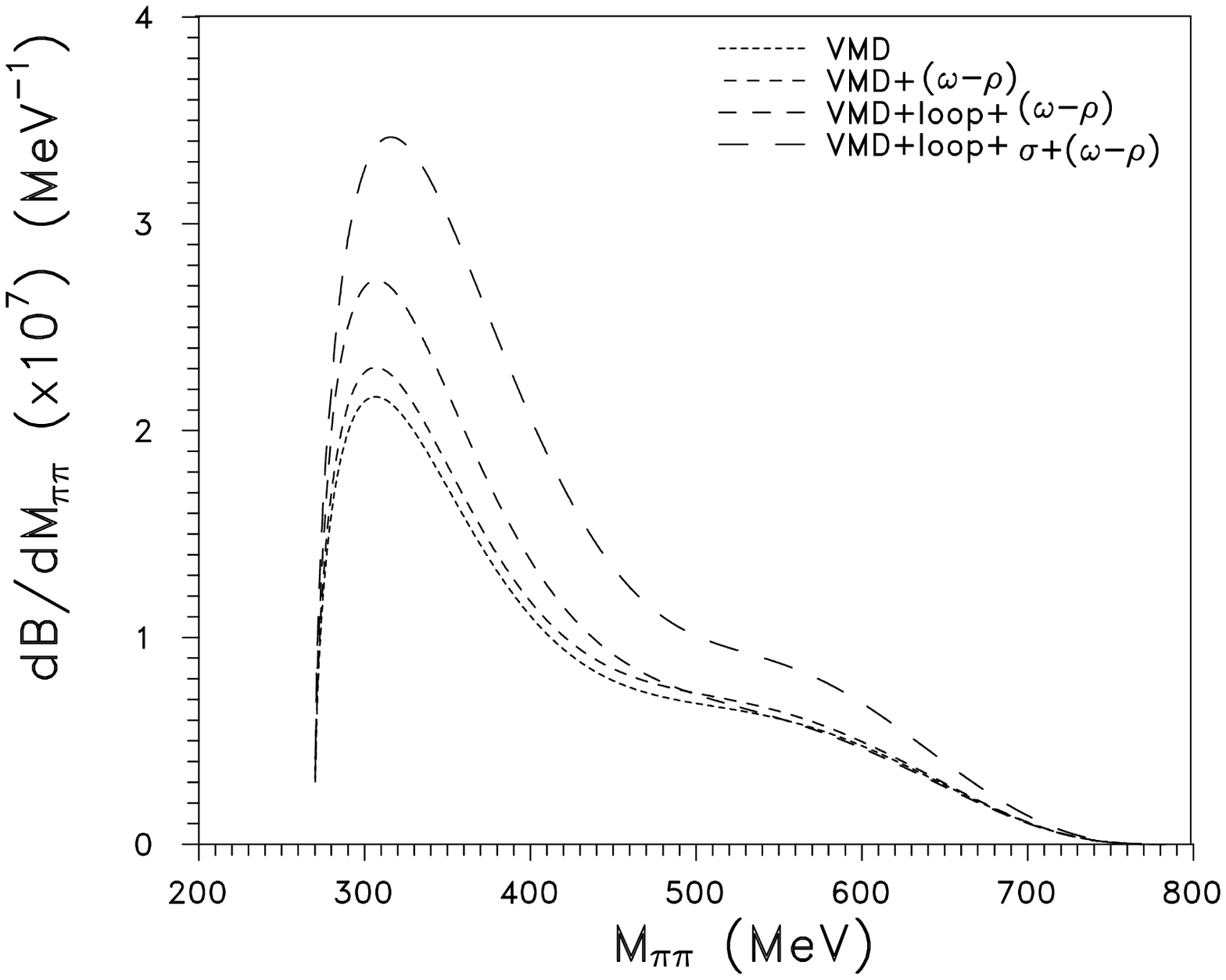,width=15cm,height=20cm}\vspace*{-3.0cm}
\caption{The $\pi^0\pi^0$ invariant mass spectrum of the decay
$\omega\rightarrow\pi^0\pi^0\gamma$ for
$g_{\omega\sigma\gamma}=0.11$. The separate contributions
resulting from the amplitudes of VMD; VMD and $\omega-\rho$
mixing; VMD, chiral loop, $\omega-\rho$ mixing; VMD, chiral loop,
$\sigma$-meson intermediate state, $\omega-\rho$ mixing are shown.
} \label{fig3}
\end{figure}

\begin{figure}
\epsfig{figure=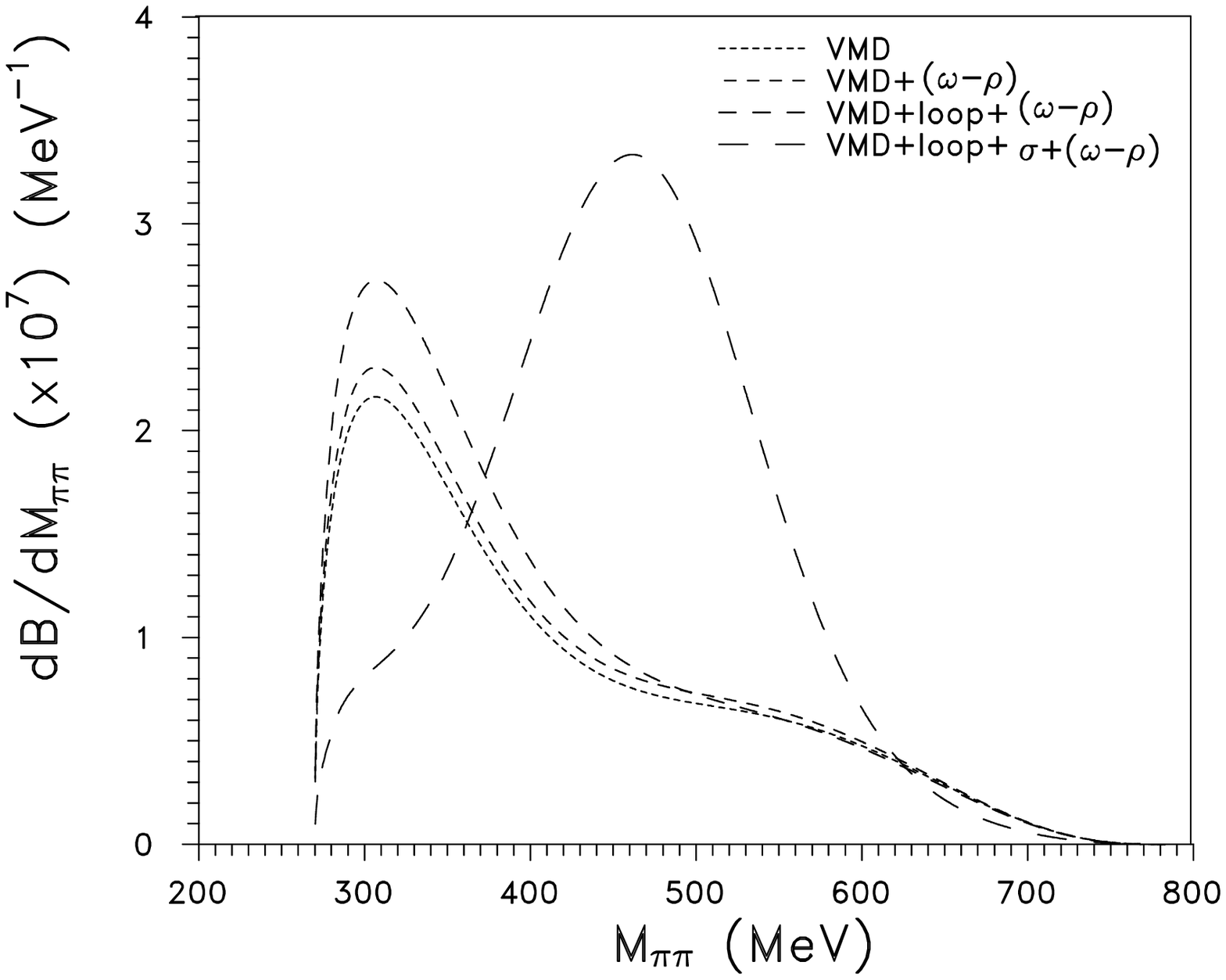,width=15cm,height=20cm} \vspace*{-3.0cm}
\caption{The $\pi^0\pi^0$ invariant mass spectrum of the decay
$\omega\rightarrow\pi^{0}\pi^0\gamma$ for
$g_{\omega\sigma\gamma}=-0.21$. The separate contributions
resulting from the amplitudes of VMD; VMD and $\omega-\rho$
mixing; VMD, chiral loop, $\omega-\rho$ mixing; VMD, chiral loop,
$\sigma$-meson intermediate state, $\omega-\rho$ mixing are shown.
} \label{fig4}
\end{figure}

\end{document}